\documentclass[twocolumn]{bmcart}

\usepackage{amsthm,amsmath}
\RequirePackage{hyperref}
\usepackage[utf8]{inputenc} 

\startlocaldefs
\usepackage{graphicx}
\graphicspath{{images/}}
\endlocaldefs

\begin{document}

\begin{frontmatter}

\begin{fmbox}
\dochead{Research}


\title{A Cloud-Fog Computing Architecture for Real-Time Digital Twins}


\author[
  addressref={aff1},                   
  corref={aff1},                       
  email={francisco.knebel@inf.ufrgs.br}   
]{\inits{F.P.K.}\fnm{Francisco P.} \snm{Knebel}}
\author[
  addressref={aff1},
  email={jwickboldt@inf.ufrgs.br}
]{\inits{J.A.W.}\fnm{Juliano A.} \snm{Wickboldt}}
\author[
  addressref={aff1},
  email={Edison Pignaton de Freitas}
]{\inits{E.P.F.}\fnm{Edison P.} \snm{de Freitas}}


\address[id=aff1]{
  \orgdiv{Institute of Informatics},             
  \orgname{Federal University of Rio Grande do Sul},          
  \city{Porto Alegre},                              
  \cny{BR}                                    
}





\begin{abstractbox}

\begin{abstract} 
Digital Twin systems are designed as two interconnected mirrored spaces, one real and one virtual, each reflecting the other, sharing information, and making predictions based on analysis and simulations. The correct behavior of a real-time Digital Twin depends not only on the logical results of computation but also on the timing constraints. To cope with the large amounts of data that need to be stored and analyzed, modern large scale Digital Twin deployments often rely on cloud-based architectures. A significant portion of the overall response time of a Digital Twin is spent on moving data from the edge to the cloud. Therefore, implementing Digital Twins using cloud-fog architectures emerges as an alternative to bring computing power closer to the edge, reducing latency and allowing faster response times. This paper studies how suitable the use of a cloud-fog architecture is to handle the real-time requirements of Digital Twins. Based on a realistic implementation and deployment of Digital Twin software components, it is possible to conclude that the distribution of Digital Twins in a fog computing setup can reduce response times, meeting its real-time application requirements.
\end{abstract}



\end{abstractbox}
\end{fmbox}

\end{frontmatter}



\section{Introduction}
\label{S:1}

The Digital Twin concept was first introduced in 2002 as a model of a system containing two interconnected mirrored spaces, one real and one virtual, each space reflecting the other while sharing real-time information~\cite{grieves:2016}. Digital Twins must be designed as real-time systems so that the virtual part can reflect the real one accurately. The correctness of the system behavior depends not only on the logical results but also on the timing constraints defined for this system~\cite{kopetz:2011}.

Essentially, a Digital Twin is a computer program that acquires real-world data about a physical system and generates simulations on how that system will be affected by this data. Long before the terminology, Digital Twins were used in action by NASA, in order to monitor and maintain machines in outer space, a situation where hands-on contact with devices would be impossible, allowing for remote diagnosis and fixing problems that are could occur ~\cite{grieves:2017}. Digital Twins can be defined as model-based representations of real-time systems, as long as they are adaptable, non-static, systems, which evolve over time with their environment \cite{schneider:2018}.

To reflect their physical counterparts, Digital Twins are composed by a large number of sensors, responsible for acquiring the used data. The rapid expansion of sensors powered by the Internet of Things (IoT) is what makes Digital Twins possible, but with this increased data generation, centralized applications oftentimes experience high latency and poor connectivity situations which affects the performance of applications. With the advent of IoT devices and its increased data volume in cloud systems, moving all data to the cloud does not satisfy time-sensitive application requirements \cite{yousefpour:2019}. Fog computing enables computing, storage, networking, and data management on network nodes in closer proximity to edge devices, allowing the processing to happen not only in the cloud, but also between the edge and the cloud \cite{yousefpour:2019}, besides on the edge \cite{7488250}.

Applications in several areas can be improved with the usage of fog computing, including smart industries \cite{khan:2020, wan:2018}, homes \cite{verma:2018}, cities \cite{tang:2015, perera:2017} and healthcare \cite{nikoloudakis:2017}, all real-time systems with time-critical processing to avoid accidents and to provide a better and more reliable experience without suffering from cloud latency. Integrating fog computing with IoT and Digital Twins enables mirroring physical and virtual spaces while achieving the necessary timing requirements of real-time systems.

Among the many challenges of implementing a fog computing network, computation offloading and data preprocessing are critical in terms of scalability and guarantee of real-time requirements \cite{li:2018}. Balancing the computational workload between multiple fog nodes, closer to the edge devices, enables the network to be more efficient in handling timing critical tasks, by distributing operations between different devices and reducing communications between edge and cloud. For Digital Twins operate under time-sensitive conditions, research is still required so that a number of connected services and devices can scale and correctly execute real-time operations in timely fashion. It is still challenging to manage, process, and store all this data in a highly distributed environment~\cite{dizdarevic:2019, khan:2020}.

Literature related to the usage of cloud and fog in IoT-related themes is already existent and discussed, but are limited to abstract and conceptual proposals for these architectures, lacking on practical demonstrations, which show absolute differences between the advantages of each deployment scenario.
In light of these challenges and possibilities, this paper proposes the usage of a cloud-fog architecture that can manage a large scale number of real-time Digital Twins, increasing performance while reducing latency, to meet timing requirements. Thus, the contribution of this work is the study of an distributed architecture to support Digital Twin applications that must meet real-time requirements, avoiding timing problems that usually happen in centralized approaches.

The remainder of this paper is structured as follows. Section \ref{S:2} provides background information for Digital Twin and cloud, fog, and edge computing concepts. Section \ref{S:3} discusses related work. Section \ref{S:4} describes the proposed solution. Section \ref{S:5} presents the experiments, the methodology, and the obtained results. Section \ref{S:6} discusses the results obtained in the experimental scenarios. Section \ref{S:7} concludes the paper and presents directions for future work.

\section{Background}
\label{S:2}

\subsection{Digital Twins}

Digital Twins are systems designed to work with rich data provided by both the insight of the models and implicit knowledge of objects, products of their behavior, only requiring the availability of raw data about their physical counterparts. To achieve this, Digital Twins must accurately simulate, analyze, and predict real-world events and situations, via the collection of real-time physical data, mirroring it into its virtual space. Digital Twins can not only anticipate and predict but can issue proactive measures to avoid problems by combining real-time data with the knowledge of other Digital Twins instances. Simulations can forecast the outcomes, allowing human operators to choose, knowing the consequences of the possible actions, and set the optimal solution~\cite{voell:2018}.

The implementation of self-aware Digital Twins creates modern control systems that adapt to their environment, not only mimicking but understanding the reasons behind the behavior of their physical counterparts~\cite{stojanovic:2018}. The virtual part of a Digital Twin system is not necessarily physically implemented along with the native physical devices. Conversely, it can be implemented physically apart from its real counterpart in a distributed and evolving manner, dynamically learning patterns, and helping predict possible disruptions throughout the life of the system~\cite{cardin:2019}.

\subsection{Cloud, Fog, and Edge Computing}

Cloud computing enables convenient, on-demand access to a shared pool of computing resources~\cite{mell:2011}. Cloud computing transfers the responsibility of data processing and storage to remote data centers, via the Internet. It allows resources to be geographically distributed, increasing data redundancy and reliability~\cite{marinescu:2017, deng:2010}. 
Fog nodes are distributed computing entities, formed by at least one networked physical processing device, able to execute distributed134 tasks~\cite{tordera:2016,reyna:2018}.
Edge computing moves processing nearest to where it is needed, allowing computation closer to the source, reducing cloud traffic and service latency, improving response times~\cite{chen:2019}.

Fog differentiates itself from the cloud due to its nodes being deployed in large numbers at less centralized locations, compared to centralized cloud data centers. In essence, fog computing is an extension of cloud computing into the physical world and their multi-connected devices, ensuring data availability where and when it is needed~\cite{atlam:2018}, while edge computing represents the nearest the processing resources can be placed regarding the data source. Positioning of the network end-devices in near proximity of the user or application increases the response speed, via minimizing network latency~\cite{ray:2019, aazam:2018}.

This work presents the idea of using both cloud and fog layers, but excluding edge. Since the edge is composed mostly of devices without complete control from the cloud provider, concerns are presented in regards to system reliability and availability. It can be used to reduce latency and enable more computing power, but due to its possible short and temporary availability, using a more controlled layer, in this case, the fog enables the objective of latency reduction while still being supervisable.

\section{Related Work}
\label{S:3}

Many studies combining cyber-physical systems, Digital Twins, and fog computing exist, but to the best of our knowledge, not in the context of real-time systems.
Cyber-physical systems using Digital Twins in the cloud, fog, or edge environments have been proposed by Qi~\textit{et al.}~\cite{qinglin:2018} and Kim \cite{kim:2019} as three levels of systems: unit, system, and system of systems (or service) level. The proposal by Qi~\textit{et al.} \cite{qinglin:2018} highlights that cloud computing enables on-demand resources and computational sharing, while fog computing shifts computation, storage, and networking of the cloud to the edge network, and edge computing allows processing at closer proximity to the data source, which can benefit cyber-physical systems. 
Kim~\cite{kim:2019} states that the key driver to moving towards edge/fog computing is time-sensitive communication, which is a required feature for real-time systems. It also mentions that distinctions between the edge, fog, and cloud layer need to be further clarified to be better explored in different setups.

A decentralized model for an edge-fog network with IoT devices is proposed by Mohan and Kangasharju~\cite{mohan:2016}. The proposal has the edge built with voluntary and loosely coupled general-purpose devices, one or two hops from the IoT resources, and horizontal connection between the edge devices. The fog layer, in this model, is a consolidation of networking devices (\textit{i.e.}, routers, switches) where edge devices can offload computationally intensive tasks. The model includes a cloud component, but it is used only for centralized data storage, with no computational capabilities. This approach would not suffice in a scenario with high computational requirements, since it has no fallback to the cloud, in cases where the edge and fog are overloaded with tasks.

The benefits of fog and edge computing, which include low latency, locality, and scalability, are well known. However, Ahmad and Afzal~\cite{ahmad:2019} state that although there are solutions for this domain, many aspects are still unexplored in practical scenarios, lacking an architecture that fully implements these principles. 
Despite this landscape, there are examples of real situations where fog computing was used to build cyber-physical systems~\cite{fernandez:2018}, to accelerate production processes. In the scenario described by Fernández-Caramés \textit{et al.}~\cite{fernandez:2018}, experiments revealed that fog nodes executed their tasks faster than the same tasks in a purely cloud-dependent environment. The authors report that these nodes can process more data while under high-load situations, in which real-time processing is constrained.

In short, given the studies presented above, currently, the Digital Twin literature still has many unexplored study areas, lacking on an actual deployment of a real implementation of Digital Twins which utilizes the benefits of a cloud-fog architecture, presenting how it was built, its components and with experiment results proving its benefits.

\section{Proposal}
\label{S:4}

This section initially describes in detail the problem addressed in this work. Then, it provides an overview of how the Digital Twin architecture was designed, its components, and details on how each of the components is used in the experiments presented in Section~\ref{S:5}.

\subsection{Addressed Problem}

The literature review revealed that the use of cloud and fog computing to build Digital Twins is not an entirely new idea. However, the studies surveyed presented mainly abstract concepts, establishing the overall idea with notions on what to expect from a cloud-based Digital Twin deployment, but without realistic implementations. An actual implementation and extensive experimentation would allow a direct analysis of how systems like this can benefit from the use of cloud, fog, and edge interchangeably, and how this could affect the performance of this network of systems.

The three layers edge-fog-cloud architecture has been presented in multiple investigations \cite{qinglin:2018, kim:2019}, but none of which presents this architecture in the context of real-time systems or analyzing how much each layer could add or reduce latency. Another problem present in this situation is the scenario of a public cloud/fog: the network resources are shared between heterogeneous systems. Systems in this network require resource competition management, defining process priority for real-time systems to execute their operations without being frozen because other systems have access to the network resources. 
The modification of existing systems to support Digital Twins implies the addition of a new real-time approach, which would share resources of a network that could already be near its computational limits.

Non-shared cloud and fog Digital Twin network scenarios presented in other works do not accurately describe all real situations, because they demonstrate scenarios where companies would build an entire infrastructure with the sole purpose of supporting Digital Twins. Real situations where they would be implemented most likely already have a working network where the Digital Twin would be inserted, requiring minimal physical change. Cloud and fog, by definition, enable reusable and cheap computation by handling multiple different and independent devices. Networks with a single purpose do not make sense because idle resources would be wasted, going against the whole concept of cloud computing.

These problems imply two situations: 1) the usage of cloud, fog, and edge computing in the context of workload variability; and 2) the definition of priority for different devices in the network to allow priority computation to be allocated to time-critical systems.
Sharing resources in a network requires the definition of a priority policy between devices, to allocate the necessary computation to Digital Twins or other systems. In light of this landscape, the problem addressed in this paper is the investigation of how appropriate a cloud-fog architecture is to support real-time Digital Twins.  

\subsection{Digital Twin Design Overview}

Modern systems built with Digital Twins in mind can implement business logic, ranging from simple computing software, only storing and processing data, to complex systems that can ``think'' and provide advanced insight for its users. The use of artificial intelligence techniques in all types of tasks is the next step in digitalization and process automation, and the use of cloud-based technologies is an essential tool for building scalable systems for any domain.

For Digital Twins to accurately simulate, analyze, and predict events and situations, they must collect a variety of data from the physical mirrored device or system and process it effectively and timely. Digital Twins must support a collection of different models to accurately describe devices in their full life-cycle phases, which may include simulation and predictive models.
As so, Digital Twins require to be equipped with a collection of services to effectively monitor and simulate the physical world it mirrors and computes relevant decisions.

The architecture we designed and implemented is an open-source\footnote{Source code available at \url{https://github.com/Open-Digital-Twin}.} software built as a collection of individual microservices, available under the GNU General Public License v3.0.
For the experiments performed in this work, the architecture contains microservices for the client devices, simulating the physical parts, a Digital Twin instance, which is the digital counterpart, and a message broker to intermediate the communication. This work focuses on the communication aspect of the Digital Twin and will not detail internal processing services. These elements are the minimum components required for the Digital Twin communication, emulating the physical and digital parts, from the model proposed by Damjanovic-Behrendt\cite{damjanovic-behrendt:2019}.

To execute the experiments, the client devices were simulated with a custom program that sends sequentially numbered messages, with a configurable payload and publication interval, to an MQTT broker, which uses a topic-based publish/subscribe pattern for communication. This allows decoupling from publisher and subscriber, allowing for each service to run independently and with higher scalability. Each client device publishes their messages with a specific topic known by the subscribed twin instance, identifying their publications as being from that specific client.
Numbered messages allow us to analyze the timing aspects of each message, from its publication by the client until it is processed by the twin instance.

The Digital Twin instance is a prototype of the digital part of the system, which implements the required mechanisms for acquisition, storage, and processing of data obtained from its physical counterpart, and is where the physical product is linked throughout its entire life-cycle~\cite{grieves:2017}. The Digital Twin instance is the final destination for the client messages after they are distributed by the broker. In this work, messages sent in the opposite direction, \textit{i.e.} from the twin instance to real devices, were not experimented with and require analysis by future work.

The current implementation uses the Eclipse \linebreak Mosquitto message broker, version 1.6.10, as it is an open-source message broker implementation of the MQTT protocol, purposely built to be open, simple, and light weight, characteristics heavily associated with IoT contexts~\cite{mqtt:2019}. Eclipse Mosquitto is intended for use in situations where there is a need for lightweight messaging, particularly on constrained devices with limited resources~\cite{mosquitto:2017}.
The broker was configured with its default settings, except for the max message queue limitations which were removed, since the broker would discard all messages received if this limit was reached, only retaining the physical restriction of memory limitations of where the broker is being executed.

To meet real-time requirements, the fact that the Digital Twin components of the architecture we designed and implemented are microservice-based allows for a variety of distribution scenarios. These components can be distributed between cloud, fog, and edge computing layers, depending on the needs of the corresponding physical system and the available network resources.

\section{Experimental Setup and Methodology}
\label{S:5}

This section describes the performed experiments. It also describes the primary experiment factors, variables that affect the experiment response, which directly impacts the performance of the proposal in the defined scenarios. Factors will be tested in different scenarios, to analyze their impact on the architecture.

\subsection{Experiment Scenarios}

\begin{figure}[ht]
\includegraphics[width=7.5cm]{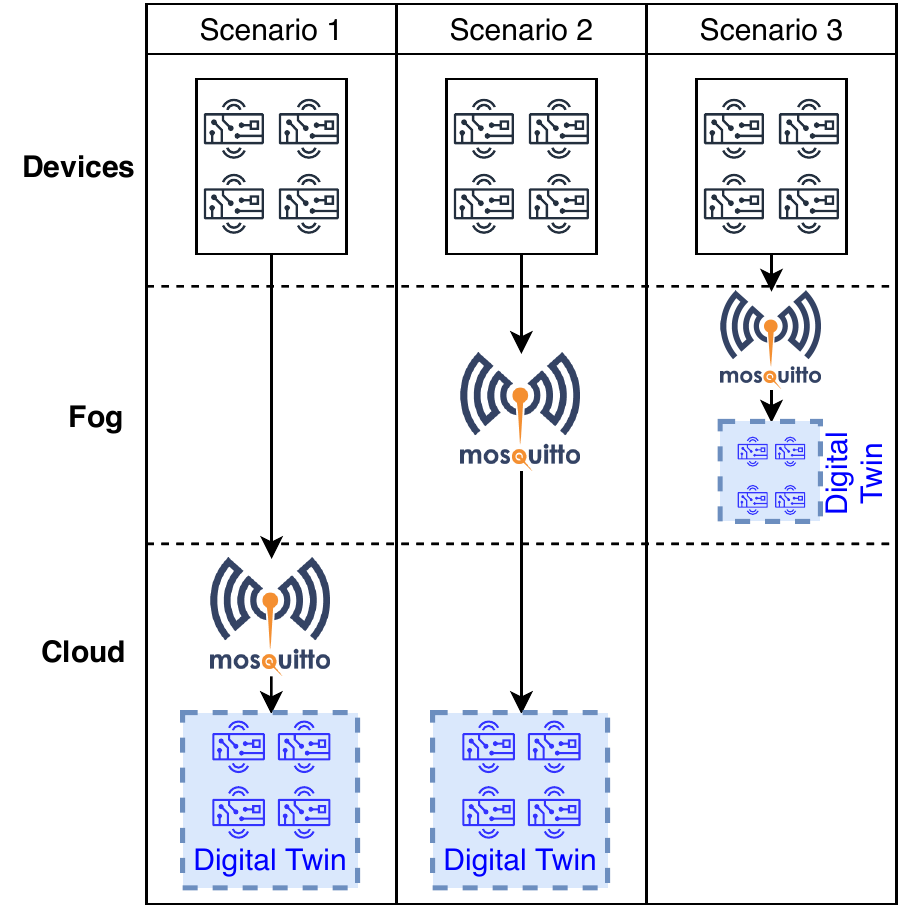}
\caption{Illustration of the three different experiment scenarios, with their corresponding distributions.}
\label{image:scenario1}
\end{figure}

To perform the proposed study, three different experiment scenarios are proposed. These three scenarios are illustrated in Figure \ref{image:scenario1} and detailed as follows:

\begin{itemize}
\item \emph{Scenario 1 - Cloud only:} 
Processing is done exclusively on a single cloud node. The MQTT broker, where the client publishes and the twin instance subscribes to a data source, resides on this cloud node.

\item \emph{Scenario 2 - Fog and Cloud:}
The MQTT broker is located on a fog node, closer to the edge, while the twin instance resides on the cloud.

\item \emph{Scenario 3 - Fog only:}
Both broker and the twin instance resides on a fog node.
\end{itemize}

\subsection{Experiment Factors}

\begin{itemize}
    \item \textit{Number of data sources:} twin elements can have multiple data sources, in the form of multiple individual devices, which by themselves can have multiple sensors. For simplification of the problem in this paper, an element is defined as one general part or device of the physical system, which has only one data source. To emulate multiple data sources, it is possible to increase the number of elements accordingly. Each source then sends a certain amount of messages to the broker.
    
    \item \textit{Message payload size:} the number of bytes sent in the MQTT message payload generated by the client devices, which will affect bandwidth usage and processing time for the message broker and the twin instance.
    
    \item \textit{Source message frequency:} the estimated time between messages sent by the data sources, or client devices, depending on the types of sensors that define the physical system.
    
    \item \textit{Transmission latency:} time for the message to travel in two scenarios: 1) from the client to the broker; and 2) from the broker to the twin instance. This latency is directly affected by the distribution of the architecture between cloud and fog.
\end{itemize}

Each scenario was individually tested under three situations: (A) varying the number of messages sent by each data source; (B) varying the frequency of messages transmissions; and (C) varying the payload of the sent messages.

\begin{itemize}
\item \emph{Situation A - Source and number of message:}
Situation A varied the amount of messages sent by each data source, in sets of 100, 1000, 10000, and 100000 messages by each source. To test concurrency, the number of data sources was set to 1, 3, and 5, which would add up to a maximum of 500000 messages to be handled by the broker and the Digital Twin instance. The delay between when each message was sent by the data sources was fixed to 10ms, with a payload of 64 bytes.

\item \emph{Situation B - Message frequency:}
Situation B varied the wait interval between each message sent by the data source, in sets of 10, 20, 40, 80, 160, 320, 640, 1280, and 2560 milliseconds. In this situation, a total of 1000 messages were sent by only one source, each with a payload of 64 bytes.

\item \emph{Situation C - Message payload size:}
Situation C tests variations of payload sizes on the messages sent by the data source. The payloads were set to 8, 16, 32, 64, 128, 256, and 512 bytes. Each test sent a total of 1000 messages from one source, with a fixed delay of 80ms between each message.
\end{itemize}

\subsection{Experiment Hardware Specifications}

A server running an Intel Xeon E5-2420 (with 6 cores and 12 threads) and 32 GB of RAM was used as the experiment environment.
The experiments were executed on three individual Kernel-based Virtual Machines simulating real devices, each named according to their use on the defined scenarios: Cloud, Fog, and Client.
The Cloud virtual machine runs with two virtual CPUs and 4GB of memory. The Fog and Client are by definition devices with more constrained resource capacities, so they only have one virtual CPU each, with 1GB of memory.

Regarding network limitations, the only limit inserted was a simulated network latency in the communication between the machines, varying according to each scenario, injected by the usage of \textit{tc-netem} \cite{netem:2011}.
In scenario 1, a delay of 100ms, with a 10ms variation, was added in the Client. In scenarios 2 and 3, two different delays were added: 40ms, varying 10ms, in the Client, and 20ms, varying 5ms, in the Fog.
Delays were injected into the scenarios to simulate the different delays when devices communicate with fog and cloud components, not only with fog having lower hardware capabilities but also having a lower delay.
Considering just two machines, without any communication delay difference, the one with more computational power would have a faster response, but computing in the edge/fog layers is the concept that communication can be the bottleneck in how specific distributed systems function.

\section{Results}
\label{S:6}

Bringing processing closer to the edge, to handle message exchange, reduces latency for the application.
Digital Twins, being composed with a multitude of distributed sensors, could experience high latency and poor connectivity. Proposing processing being done closer to edge is a possible strategy, since being purely dependent on the cloud may not satisfy time-sensitive requirements \cite{yousefpour:2019}.
The processing unit can receive the data from the source devices and compute its suggestion to the physical system with a shorter response time.

\begin{figure}[ht]
\includegraphics[width=7.5cm]{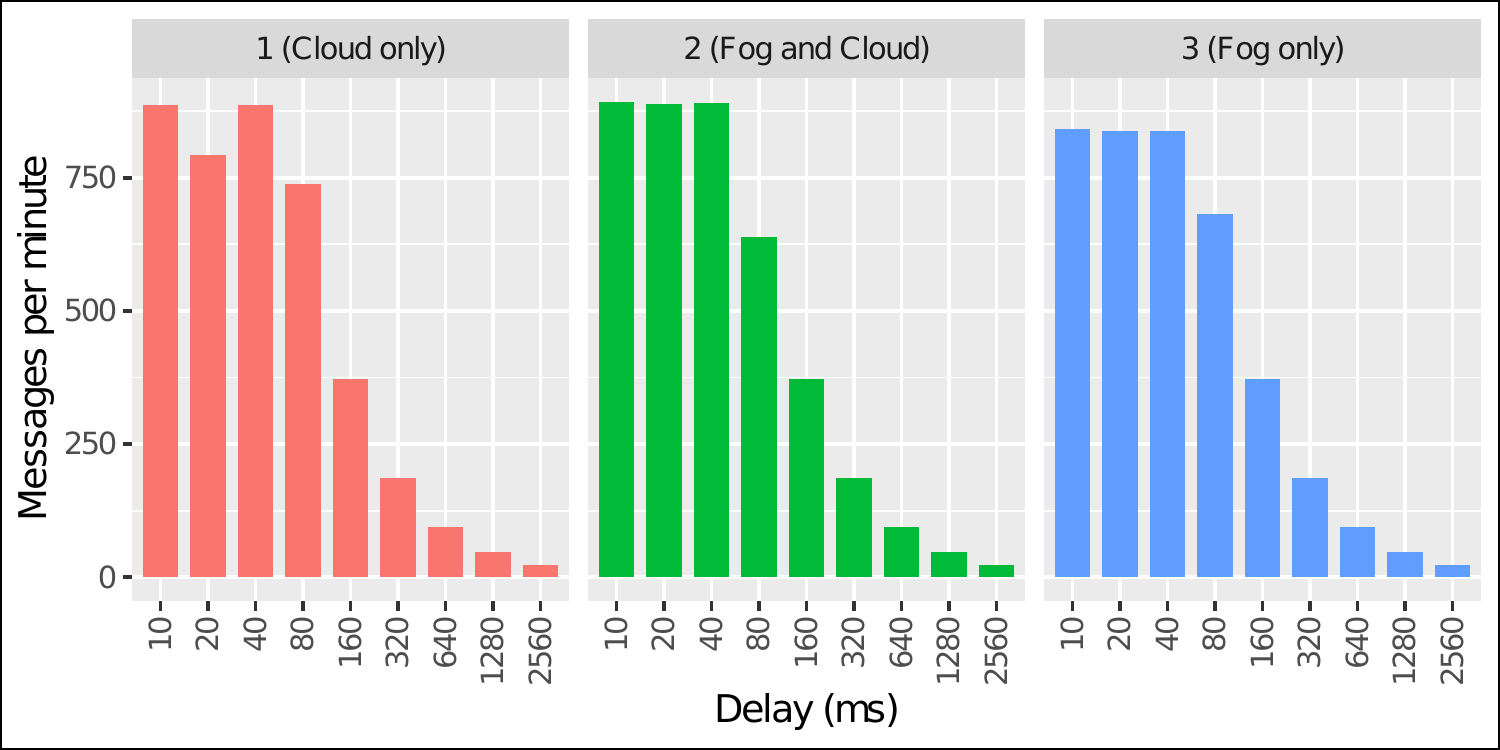}
\caption{Messages per minute received by the Digital Twin instance, in each scenario, in the message frequency situation.}
\label{delay-experiment}
\end{figure}

An important observation about how the MQTT broker throttles messages was found when testing Situation B, varying the message frequency, in the described scenarios. As evidenced by Figure~\ref{delay-experiment}, which shows the amount of messages received per minute in each scenario, varying the delay between publications, the broker has a limit on the number of messages it can receive and publish to its subscribers in the order of 800 messages per minute. We did not set a limit to the message queue in order to prevent the broker from dropping messages under high load situations. 
Thus, this message per minute rate is a hard limit set by the broker in order to prioritize receiving publications and to not overload subscribers through limiting the number of in-flight messages. Eclipse Mosquitto can be configured to increase this in-flight message limitation or even remove it. However, doing so could cause message ordering to break or it could overload subscribers and should be tested in a case by case scenario, according to the number of messages sent by the real device.

\begin{figure}[ht]
\includegraphics[width=7.5cm]{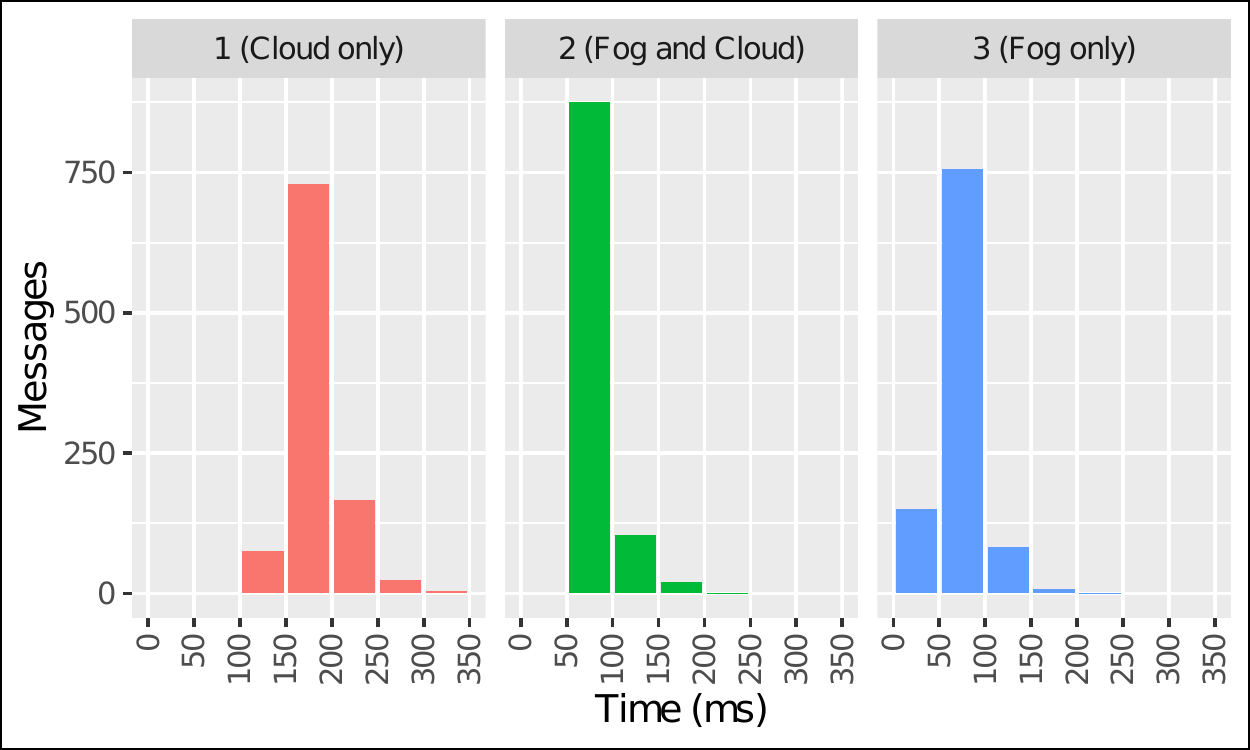}
\caption{Timing histogram of 1000 messages, from client publish until reaching the Digital Twin, in each scenario.}
\label{message-experiment}
\end{figure}

Figure~\ref{message-experiment} presents a timing histogram of 1000 messages in the three scenarios, on situation C with a payload of 64 bytes and 80ms interval between each message, from the time it is published by the client until received by the Digital Twin instance (subscriber). Results from this experiment show not just the average timing of messages, but the time of each individual publication. Scenario 1 presents messages with a higher time compared to the two other scenarios, where most messages are received in the 50 to 100 ms interval, in this experiment. Considering a real-time system with a communication requirement of 200ms in the Cloud only scenario (1), 805 of the 1000 messages would respect this limitation, representing 80.5\% of the total messages sent by the client. Meanwhile, only one message would not arrive in time for this imposed system limitation in the Fog and Cloud (2) and Fog only (3) scenarios, respectively.
Further reducing this limitation to 150ms, 92.4\% of messages sent in scenario 1 would be lost, while 0.21\% and 0.09\% would not respect the requirement by scenarios 2 and 3.

\begin{figure*}[ht!]
\includegraphics[width=16.5cm]{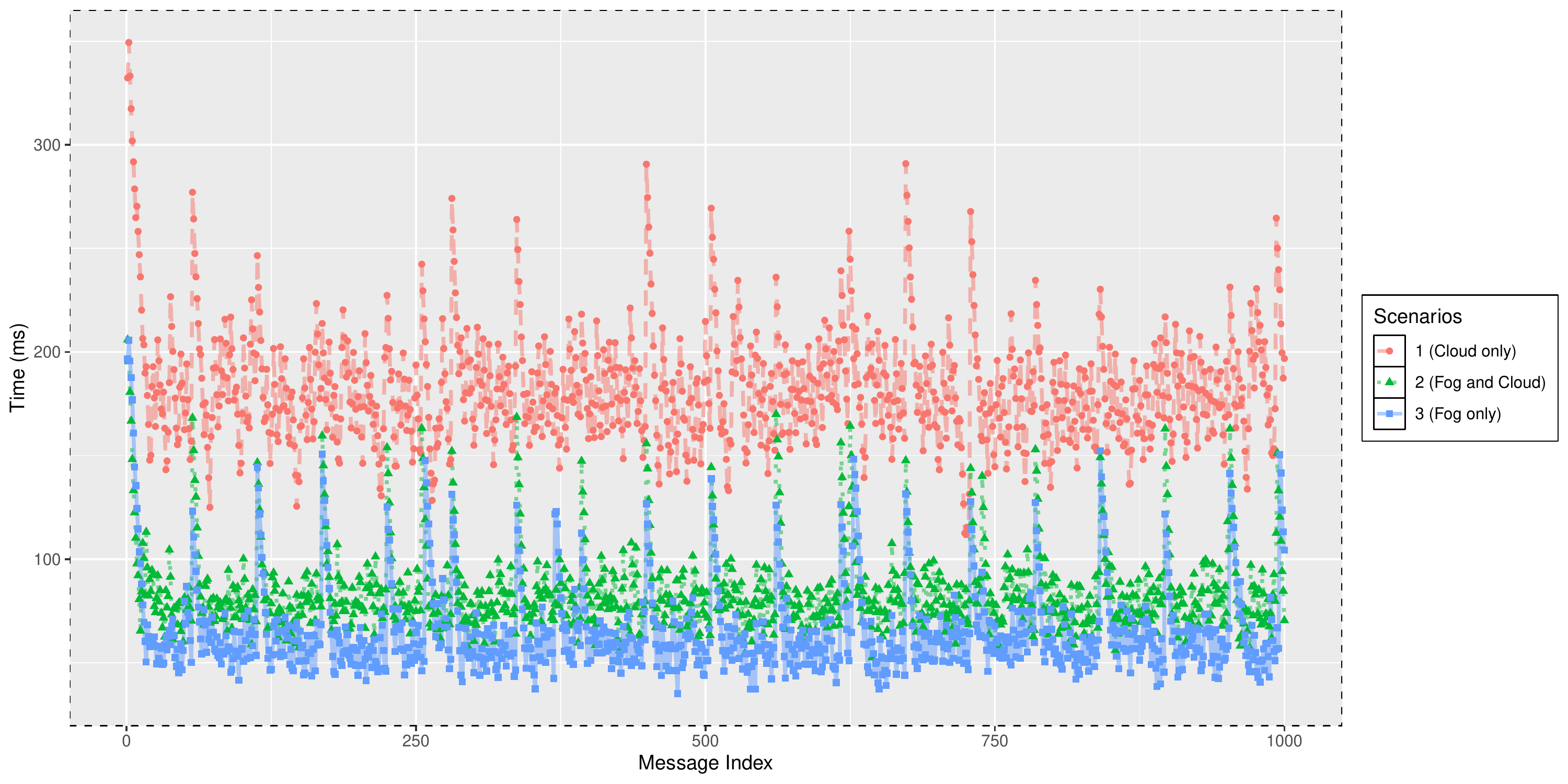}
\caption{Timing per message, from client publish to the Digital Twin receiving, in each scenario, sending 1000 messages with a 80ms delay between each publication.}
\label{delay-80-experiment}
\end{figure*}

More benefits depend on the architecture scale of implementation in the fog/cloud levels. By having resources closer to the edge, it is possible to perform preprocessing as soon as possible, before centralizing the data in the cloud. Figure~\ref{delay-80-experiment} makes this clear by showing the time each message takes from being sent by the client until received by the twin instance in each deployment scenario. The situation shown in this image sends 1000 messages, each one with a delay between each publication of 80ms. Table~\ref{delay-80-experiment-table} provides a summation of the average times of this experiment, providing a general overview of all the published messages.

\begin{table}[!htb]
\caption{Average times per message, in milliseconds, from client to twin, for the 80ms delay experiment situation.}
\label{delay-80-experiment-table}
\centering
\begin{tabular}{|c|c|c|c|c|}
\hline
Scenario & Average & Std Deviation & Max & Min \\
\hline
1 & 182.45 & 27.932 & 349.35 & 111.82 \\
\hline
2 & 84.43 & 20.417 & 205.83 & 52.51 \\
\hline
3 & 66.08 & 22.626 & 205.63 & 35.13 \\
\hline
\end{tabular}
\end{table}

The usage of cloud computing enables devices to store data in a centralized location, but large amounts of data being transmitted, natural to the concept of Digital Twins with a large scale network of sensors, implies an increasing delay, affecting response times from the server.

As shown in Table~\ref{delay-80-experiment-table}, deploying to the Fog and Cloud scenario (2) and the Fog only scenario (3) reduced average times per message by 54\% and 64\% respectively, compared to the Cloud only scenario (1). Maximum times were also reduced by 41\% for scenarios 2 and 3, as were the minimum times by 53\% and 69\%.

As a response from the results presented in this work, fog computing is shown as a viable alternative to reduce the effects of this problem, reducing the transmission delay in order to match real-time requirements.
Further reductions depend on improving lower layers of the network components, via extra connectivity and reduced latency in general.

\section{Conclusion and future work}
\label{S:7}

This paper investigated how the usage of a cloud-fog architecture for Digital Twins can help in meeting real-time requirements of these systems.

The tested architecture includes a client, for sending sensor data, the MQTT broker, for handling messages sent by the client, and the Digital Twin instance, which is subscribed by the broker. Tests were executed in three different scenarios, changing the deployment locations of the architectural elements where the broker and twin reside in the cloud, where the broker resides in a fog node, and the twin on the cloud, and lastly where both the broker and twin share the same fog node.

The experiments tested these different scenarios on how they would handle the number of messages, different frequency of messages sent, and payload size. From these experiments, it was concluded that the distribution of these Digital Twin elements closer to the edge reduce communication delays for the end application, allowing faster response by bringing preprocessing into distributed locations. This particular distribution allows the Digital Twin to meet real-time systems timing specifications.

Having all the required data flow from the client devices, the Digital Twin can run application-specific algorithms handling this input and be able to suggest actions to the physical systems respecting their real-time requirements. 

As future work, some limitations could be addressed, as the problem regarding the real-time priority policy for the MQTT broker: message handling should prioritize real-time subscribers. The MQTT protocol has no definition of priority for subscriptions, but a broker-defined configuration indicating priority topics could be used, dedicating more resources to attempt a best-effort approach into meeting real-time requirements.


\begin{backmatter}
    
\section*{Abbreviations}
IoT: Internet of Things; MQTT: Message Queuing Telemetry Transport.
    
\section*{Availability of data and materials}
Please contact authors for data requests.
    
\section*{Competing interests}
The authors declare that they have no competing interests.

\section*{Funding}
This work was supported in part by the Brazilian National Council for Scientific and Technological Development (CNPq) and in part by the Coordenação de Aperfeiçoamento de Pessoal de Nível Superior - Brasil (CAPES) - Finance Code 001.
We also thank the funding of CNPq, Research Productivity Scholarship grants ref. 313893/2018-7, ref. 306614/2017-0, ref. 309505/2020-8 and ref. 420109/2018-8.

\section*{Acknowledgements}
Not applicable.

\section*{Authors' contributions}
F. Knebel collaborated with the problem formalization and metholodogy, partipated in the software implementation and resulting data investigation.
J. Wickboldt collaborated with the problem formalization and metholodogy, partipated in the software implementation and resulting data investigation.
E. de Freitas collaborated with the problem formalization and metholodogy, and the result investigation.
All authors wrote, read, and approved the final manuscript.


\bibliographystyle{bmc-mathphys} 
\bibliography{bmc_article}      


\begin{thebibliography}{35}
\ifx \bisbn   \undefined \def \bisbn  #1{ISBN #1}\fi
\ifx \binits  \undefined \def \binits#1{#1}\fi
\ifx \bauthor  \undefined \def \bauthor#1{#1}\fi
\ifx \batitle  \undefined \def \batitle#1{#1}\fi
\ifx \bjtitle  \undefined \def \bjtitle#1{#1}\fi
\ifx \bvolume  \undefined \def \bvolume#1{\textbf{#1}}\fi
\ifx \byear  \undefined \def \byear#1{#1}\fi
\ifx \bissue  \undefined \def \bissue#1{#1}\fi
\ifx \bfpage  \undefined \def \bfpage#1{#1}\fi
\ifx \blpage  \undefined \def \blpage #1{#1}\fi
\ifx \burl  \undefined \def \burl#1{\textsf{#1}}\fi
\ifx \doiurl  \undefined \def \doiurl#1{\textsf{#1}}\fi
\ifx \betal  \undefined \def \betal{\textit{et al.}}\fi
\ifx \binstitute  \undefined \def \binstitute#1{#1}\fi
\ifx \binstitutionaled  \undefined \def \binstitutionaled#1{#1}\fi
\ifx \bctitle  \undefined \def \bctitle#1{#1}\fi
\ifx \beditor  \undefined \def \beditor#1{#1}\fi
\ifx \bpublisher  \undefined \def \bpublisher#1{#1}\fi
\ifx \bbtitle  \undefined \def \bbtitle#1{#1}\fi
\ifx \bedition  \undefined \def \bedition#1{#1}\fi
\ifx \bseriesno  \undefined \def \bseriesno#1{#1}\fi
\ifx \blocation  \undefined \def \blocation#1{#1}\fi
\ifx \bsertitle  \undefined \def \bsertitle#1{#1}\fi
\ifx \bsnm \undefined \def \bsnm#1{#1}\fi
\ifx \bsuffix \undefined \def \bsuffix#1{#1}\fi
\ifx \bparticle \undefined \def \bparticle#1{#1}\fi
\ifx \barticle \undefined \def \barticle#1{#1}\fi
\ifx \bconfdate \undefined \def \bconfdate #1{#1}\fi
\ifx \botherref \undefined \def \botherref #1{#1}\fi
\ifx \url \undefined \def \url#1{\textsf{#1}}\fi
\ifx \bchapter \undefined \def \bchapter#1{#1}\fi
\ifx \bbook \undefined \def \bbook#1{#1}\fi
\ifx \bcomment \undefined \def \bcomment#1{#1}\fi
\ifx \oauthor \undefined \def \oauthor#1{#1}\fi
\ifx \citeauthoryear \undefined \def \citeauthoryear#1{#1}\fi
\ifx \endbibitem  \undefined \def \endbibitem {}\fi
\ifx \bconflocation  \undefined \def \bconflocation#1{#1}\fi
\ifx \arxivurl  \undefined \def \arxivurl#1{\textsf{#1}}\fi
\csname PreBibitemsHook\endcsname

\bibitem{grieves:2016}
\begin{botherref}
\oauthor{\bsnm{Grieves}, \binits{M.}}:
{Origins of the Digital Twin Concept}.
Technical report,
{Florida Institute of Technology / NASA}
(2016)
\end{botherref}
\endbibitem

\bibitem{kopetz:2011}
\begin{bbook}
\bauthor{\bsnm{Kopetz}, \binits{H.}}:
\bbtitle{Real-time Systems: Design Principles for Distributed Embedded
  Applications}.
\bpublisher{Springer},
\blocation{Wien, AT}
(\byear{2011})
\end{bbook}
\endbibitem

\bibitem{grieves:2017}
\begin{bchapter}
\bauthor{\bsnm{Grieves}, \binits{M.}},
\bauthor{\bsnm{Vickers}, \binits{J.}}:
\bctitle{Digital twin: Mitigating unpredictable, undesirable emergent behavior
  in complex systems}.
In: \beditor{\bsnm{Kahlen}, \binits{F.-J.}},
\beditor{\bsnm{Flumerfelt}, \binits{S.}},
\beditor{\bsnm{Alves}, \binits{A.}} (eds.)
\bbtitle{Transdisciplinary Perspectives on Complex Systems: New Findings and
  Approaches},
\bedition{1st} edn.,
pp. \bfpage{85}--\blpage{113}.
\bpublisher{Springer},
\blocation{Rochester, USA}
(\byear{2017})
\end{bchapter}
\endbibitem

\bibitem{schneider:2018}
\begin{barticle}
\bauthor{\bsnm{Schneider}, \binits{D.}},
\bauthor{\bsnm{Trapp}, \binits{M.}}:
\batitle{{B-space: dynamic management and assurance of open systems of
  systems}}.
\bjtitle{{Journal of Internet Services and Applications}}
\bvolume{9}(\bissue{1}),
\bfpage{1}--\blpage{16}
(\byear{2018})
\end{barticle}
\endbibitem

\bibitem{yousefpour:2019}
\begin{barticle}
\bauthor{\bsnm{Yousefpour}, \binits{A.}},
\bauthor{\bsnm{Fung}, \binits{C.}},
\bauthor{\bsnm{Nguyen}, \binits{T.}},
\bauthor{\bsnm{Kadiyala}, \binits{K.}},
\bauthor{\bsnm{Jalali}, \binits{F.}},
\bauthor{\bsnm{Niakanlahiji}, \binits{A.}},
\bauthor{\bsnm{Kong}, \binits{J.}},
\bauthor{\bsnm{Jue}, \binits{J.P.}}:
\batitle{All one needs to know about fog computing and related edge computing
  paradigms: A complete survey}.
\bjtitle{Journal of Systems Architecture}
\bvolume{98},
\bfpage{289}--\blpage{330}
(\byear{2019})
\end{barticle}
\endbibitem

\bibitem{7488250}
\begin{barticle}
\bauthor{\bsnm{{Shi}}, \binits{W.}},
\bauthor{\bsnm{{Cao}}, \binits{J.}},
\bauthor{\bsnm{{Zhang}}, \binits{Q.}},
\bauthor{\bsnm{{Li}}, \binits{Y.}},
\bauthor{\bsnm{{Xu}}, \binits{L.}}:
\batitle{Edge computing: Vision and challenges}.
\bjtitle{IEEE Internet of Things Journal}
\bvolume{3}(\bissue{5}),
\bfpage{637}--\blpage{646}
(\byear{2016})
\end{barticle}
\endbibitem

\bibitem{khan:2020}
\begin{barticle}
\bauthor{\bsnm{Khan}, \binits{W.}},
\bauthor{\bsnm{Rehman}, \binits{M.}},
\bauthor{\bsnm{Zangoti}, \binits{H.}},
\bauthor{\bsnm{Afzal}, \binits{M.}},
\bauthor{\bsnm{Armi}, \binits{N.}},
\bauthor{\bsnm{Salah}, \binits{K.}}:
\batitle{Industrial internet of things: Recent advances, enabling technologies
  and open challenges}.
\bjtitle{Computers \& Electrical Engineering}
\bvolume{81},
\bfpage{106522}
(\byear{2020})
\end{barticle}
\endbibitem

\bibitem{wan:2018}
\begin{barticle}
\bauthor{\bsnm{Wan}, \binits{J.}},
\bauthor{\bsnm{Chen}, \binits{B.}},
\bauthor{\bsnm{Wang}, \binits{S.}},
\bauthor{\bsnm{Xia}, \binits{M.}},
\bauthor{\bsnm{Li}, \binits{D.}},
\bauthor{\bsnm{Liu}, \binits{C.}}:
\batitle{Fog computing for energy-aware load balancing and scheduling in smart
  factory}.
\bjtitle{IEEE Transactions on Industrial Informatics}
\bvolume{14}(\bissue{10}),
\bfpage{4548}--\blpage{4556}
(\byear{2018})
\end{barticle}
\endbibitem

\bibitem{verma:2018}
\begin{barticle}
\bauthor{\bsnm{Verma}, \binits{P.}},
\bauthor{\bsnm{Sood}, \binits{S.K.}}:
\batitle{{Fog assisted-IoT enabled patient health monitoring in smart homes}}.
\bjtitle{IEEE Internet of Things Journal}
\bvolume{5}(\bissue{3}),
\bfpage{1789}--\blpage{1796}
(\byear{2018})
\end{barticle}
\endbibitem

\bibitem{tang:2015}
\begin{bchapter}
\bauthor{\bsnm{Tang}, \binits{B.}},
\bauthor{\bsnm{Chen}, \binits{Z.}},
\bauthor{\bsnm{Hefferman}, \binits{G.}},
\bauthor{\bsnm{Wei}, \binits{T.}},
\bauthor{\bsnm{He}, \binits{H.}},
\bauthor{\bsnm{Yang}, \binits{Q.}}:
\bctitle{A hierarchical distributed fog computing architecture for big data
  analysis in smart cities}.
In: \bbtitle{Proceedings of the ASE BigData \& SocialInformatics 2015},
pp. \bfpage{1}--\blpage{6}
(\byear{2015})
\end{bchapter}
\endbibitem

\bibitem{perera:2017}
\begin{barticle}
\bauthor{\bsnm{Perera}, \binits{C.}},
\bauthor{\bsnm{Qin}, \binits{Y.}},
\bauthor{\bsnm{Estrella}, \binits{J.C.}},
\bauthor{\bsnm{Reiff-Marganiec}, \binits{S.}},
\bauthor{\bsnm{Vasilakos}, \binits{A.V.}}:
\batitle{Fog computing for sustainable smart cities: A survey}.
\bjtitle{ACM Computing Surveys (CSUR)}
\bvolume{50}(\bissue{3}),
\bfpage{1}--\blpage{43}
(\byear{2017})
\end{barticle}
\endbibitem

\bibitem{nikoloudakis:2017}
\begin{bchapter}
\bauthor{\bsnm{Nikoloudakis}, \binits{Y.}},
\bauthor{\bsnm{Markakis}, \binits{E.}},
\bauthor{\bsnm{Mastorakis}, \binits{G.}},
\bauthor{\bsnm{Pallis}, \binits{E.}},
\bauthor{\bsnm{Skianis}, \binits{C.}}:
\bctitle{{An NFV-powered emergency system for smart enhanced living
  environments}}.
In: \bbtitle{2017 IEEE Conference on Network Function Virtualization and
  Software Defined Networks (NFV-SDN)},
pp. \bfpage{258}--\blpage{263}
(\byear{2017}).
\bcomment{IEEE}
\end{bchapter}
\endbibitem

\bibitem{li:2018}
\begin{barticle}
\bauthor{\bsnm{Li}, \binits{C.}},
\bauthor{\bsnm{Xue}, \binits{Y.}},
\bauthor{\bsnm{Wang}, \binits{J.}},
\bauthor{\bsnm{Zhang}, \binits{W.}},
\bauthor{\bsnm{Li}, \binits{T.}}:
\batitle{Edge-oriented computing paradigms: A survey on architecture design and
  system management}.
\bjtitle{ACM Computing Surveys (CSUR)}
\bvolume{51}(\bissue{2}),
\bfpage{1}--\blpage{34}
(\byear{2018})
\end{barticle}
\endbibitem

\bibitem{dizdarevic:2019}
\begin{botherref}
\oauthor{\bsnm{Dizdarevi\'{c}}, \binits{J.}},
\oauthor{\bsnm{Carpio}, \binits{F.}},
\oauthor{\bsnm{Jukan}, \binits{A.}},
\oauthor{\bsnm{Masip-Bruin}, \binits{X.}}:
A survey of communication protocols for internet of things and related
  challenges of fog and cloud computing integration.
ACM Comput. Surv.
\textbf{51}(6)
(2019)
\end{botherref}
\endbibitem

\bibitem{voell:2018}
\begin{bchapter}
\bauthor{\bsnm{Voell}, \binits{C.}},
\bauthor{\bsnm{Chatterjee}, \binits{P.}},
\bauthor{\bsnm{Rauch}, \binits{A.}},
\bauthor{\bsnm{Golovatchev}, \binits{J.}}:
\bctitle{{How Digital Twins Enable the Next Level of PLM – A Guide for the
  Concept and the Implementation in the Internet of Everything Era}}.
In: \bbtitle{Product Lifecycle Management to Support Industry 4.0},
pp. \bfpage{238}--\blpage{249}.
\bpublisher{Springer},
\blocation{Cham, CH}
(\byear{2018})
\end{bchapter}
\endbibitem

\bibitem{stojanovic:2018}
\begin{bchapter}
\bauthor{\bsnm{Stojanovic}, \binits{N.}},
\bauthor{\bsnm{Milenovic}, \binits{D.}}:
\bctitle{{Data-driven Digital Twin approach for process optimization: an
  industry use case}}.
In: \bbtitle{2018 IEEE International Conference on Big Data}
(\byear{2018})
\end{bchapter}
\endbibitem

\bibitem{cardin:2019}
\begin{barticle}
\bauthor{\bsnm{Cardin}, \binits{O.}}:
\batitle{{Classification of cyber-physical production systems applications:
  Proposition of an analysis framework}}.
\bjtitle{Computers in Industry}
\bvolume{104},
\bfpage{11}--\blpage{21}
(\byear{2019}).
doi:\doiurl{10.1016/j.compind.2018.10.002}
\end{barticle}
\endbibitem

\bibitem{mell:2011}
\begin{botherref}
\oauthor{\bsnm{Mell}, \binits{P.}},
\oauthor{\bsnm{Grance}, \binits{T.}}, et al.:
{The NIST definition of cloud computing}.
National Institute of Science and Technology, Special Publication, 800
\textbf{145}
(2011)
\end{botherref}
\endbibitem

\bibitem{marinescu:2017}
\begin{bbook}
\bauthor{\bsnm{Marinescu}, \binits{D.C.}}:
\bbtitle{Cloud Computing: Theory and Practice}.
\bpublisher{Morgan Kaufmann},
\blocation{Waltham, USA}
(\byear{2017})
\end{bbook}
\endbibitem

\bibitem{deng:2010}
\begin{bchapter}
\bauthor{\bsnm{Deng}, \binits{J.}},
\bauthor{\bsnm{Huang}, \binits{S.C.-H.}},
\bauthor{\bsnm{Han}, \binits{Y.S.}},
\bauthor{\bsnm{Deng}, \binits{J.H.}}:
\bctitle{Fault-tolerant and reliable computation in cloud computing}.
In: \bbtitle{2010 IEEE Globecom Workshops},
pp. \bfpage{1601}--\blpage{1605}
(\byear{2010}).
\bcomment{IEEE}
\end{bchapter}
\endbibitem

\bibitem{tordera:2016}
\begin{botherref}
\oauthor{\bsnm{Tordera}, \binits{E.M.}},
\oauthor{\bsnm{Masip-Bruin}, \binits{X.}},
\oauthor{\bsnm{Garcia-Alminana}, \binits{J.}},
\oauthor{\bsnm{Jukan}, \binits{A.}},
\oauthor{\bsnm{Ren}, \binits{G.-J.}},
\oauthor{\bsnm{Zhu}, \binits{J.}},
\oauthor{\bsnm{Farr{\'e}}, \binits{J.}}:
What is a fog node? a tutorial on current concepts towards a common definition.
arXiv preprint arXiv:1611.09193
(2016)
\end{botherref}
\endbibitem

\bibitem{reyna:2018}
\begin{barticle}
\bauthor{\bsnm{Reyna}, \binits{A.}},
\bauthor{\bsnm{Martín}, \binits{C.}},
\bauthor{\bsnm{Chen}, \binits{J.}},
\bauthor{\bsnm{Soler}, \binits{E.}},
\bauthor{\bsnm{Díaz}, \binits{M.}}:
\batitle{On blockchain and its integration with iot. challenges and
  opportunities}.
\bjtitle{Future Generation Computer Systems}
\bvolume{88},
\bfpage{173}--\blpage{190}
(\byear{2018})
\end{barticle}
\endbibitem

\bibitem{chen:2019}
\begin{botherref}
\oauthor{\bsnm{Chen}, \binits{Y.}},
\oauthor{\bsnm{Deng}, \binits{S.}},
\oauthor{\bsnm{Ma}, \binits{H.}},
\oauthor{\bsnm{Yin}, \binits{J.}}:
Deploying data-intensive applications with multiple services components on
  edge.
Mobile Networks and Applications,
1--16
(2019)
\end{botherref}
\endbibitem

\bibitem{atlam:2018}
\begin{barticle}
\bauthor{\bsnm{Atlam}, \binits{H.F.}},
\bauthor{\bsnm{Walters}, \binits{R.J.}},
\bauthor{\bsnm{Wills}, \binits{G.B.}}:
\batitle{{Fog Computing and the Internet of Things: A Review}}.
\bjtitle{Big Data and Cognitive Computing}
\bvolume{2}(\bissue{2}),
\bfpage{10}
(\byear{2018})
\end{barticle}
\endbibitem

\bibitem{ray:2019}
\begin{barticle}
\bauthor{\bsnm{Ray}, \binits{P.P.}},
\bauthor{\bsnm{Dash}, \binits{D.}},
\bauthor{\bsnm{De}, \binits{D.}}:
\batitle{{Edge computing for Internet of Things: A survey, e-healthcare case
  study and future direction}}.
\bjtitle{{Journal of Network and Computer Applications}}
\bvolume{140},
\bfpage{1}--\blpage{22}
(\byear{2019})
\end{barticle}
\endbibitem

\bibitem{aazam:2018}
\begin{barticle}
\bauthor{\bsnm{Aazam}, \binits{M.}},
\bauthor{\bsnm{Zeadally}, \binits{S.}},
\bauthor{\bsnm{Harras}, \binits{K.A.}}:
\batitle{Offloading in fog computing for iot: Review, enabling technologies,
  and research opportunities}.
\bjtitle{Future Generation Computer Systems}
\bvolume{87},
\bfpage{278}--\blpage{289}
(\byear{2018})
\end{barticle}
\endbibitem

\bibitem{qinglin:2018}
\begin{bchapter}
\bauthor{\bsnm{Qi}, \binits{Q.}},
\bauthor{\bsnm{Zhao}, \binits{D.}},
\bauthor{\bsnm{Liao}, \binits{T.W.}},
\bauthor{\bsnm{Tao}, \binits{F.}}:
\bctitle{{Modeling of Cyber-Physical Systems and Digital Twin Based on Edge
  Computing, Fog Computing and Cloud Computing Towards Smart Manufacturing}}.
In: \bbtitle{ASME 2018 13th International Manufacturing Science and Engineering
  Conference}
(\byear{2018}).
\bcomment{American Society of Mechanical Engineers Digital Collection}
\end{bchapter}
\endbibitem

\bibitem{kim:2019}
\begin{bchapter}
\bauthor{\bsnm{Kim}, \binits{T.}}:
\bctitle{{Cyber Physical Systems Framework of Edge-, Fog-, and
  Cloud-Computing}}.
In: \bbtitle{Proceedings of the International Conference on Embedded Systems,
  Cyber-physical Systems, and Applications (ESCS)},
pp. \bfpage{10}--\blpage{14}
(\byear{2019}).
\bcomment{The Steering Committee of The World Congress in Computer Science,
  Computer Engineering and Applied Computing (WorldComp)}
\end{bchapter}
\endbibitem

\bibitem{mohan:2016}
\begin{bchapter}
\bauthor{\bsnm{Mohan}, \binits{N.}},
\bauthor{\bsnm{Kangasharju}, \binits{J.}}:
\bctitle{{Edge-Fog cloud: A distributed cloud for Internet of Things
  computations}}.
In: \bbtitle{2016 Cloudification of the Internet of Things (CIoT)},
pp. \bfpage{1}--\blpage{6}
(\byear{2016}).
\bcomment{IEEE}
\end{bchapter}
\endbibitem

\bibitem{ahmad:2019}
\begin{bchapter}
\bauthor{\bsnm{Ahmad}, \binits{S.}},
\bauthor{\bsnm{Afzal}, \binits{M.M.}}:
\bctitle{{Deployment of Fog and Edge Computing in IoT for Cyber-Physical
  Infrastructures in the 5G Era}}.
In: \bbtitle{International Conference on Sustainable Communication Networks and
  Application},
pp. \bfpage{351}--\blpage{359}
(\byear{2019}).
\bcomment{Springer}
\end{bchapter}
\endbibitem

\bibitem{fernandez:2018}
\begin{barticle}
\bauthor{\bsnm{Fern{\'a}ndez-Caram{\'e}s}, \binits{T.M.}},
\bauthor{\bsnm{Fraga-Lamas}, \binits{P.}},
\bauthor{\bsnm{Su{\'a}rez-Albela}, \binits{M.}},
\bauthor{\bsnm{D{\'\i}az-Bouza}, \binits{M.A.}}:
\batitle{{A Fog Computing Based Cyber-Physical System for the Automation of
  Pipe-Related Tasks in the Industry 4.0 Shipyard}}.
\bjtitle{Sensors}
\bvolume{18}(\bissue{6}),
\bfpage{1961}
(\byear{2018})
\end{barticle}
\endbibitem

\bibitem{damjanovic-behrendt:2019}
\begin{barticle}
\bauthor{\bsnm{Damjanovic-Behrendt}, \binits{V.}},
\bauthor{\bsnm{Behrendt}, \binits{W.}}:
\batitle{An open source approach to the design and implementation of digital
  twins for smart manufacturing}.
\bjtitle{International Journal of Computer Integrated Manufacturing}
\bvolume{32}(\bissue{4-5}),
\bfpage{366}--\blpage{384}
(\byear{2019})
\end{barticle}
\endbibitem

\bibitem{mqtt:2019}
\begin{botherref}
\oauthor{\bsnm{Banks}, \binits{A.}},
\oauthor{\bsnm{Briggs}, \binits{E.}},
\oauthor{\bsnm{Borgendale}, \binits{K.}},
\oauthor{\bsnm{Gupta}, \binits{R.}}:
{MQTT Version 5.0}.
Accessed: 2020-08-09
(2019).
\url{https://docs.oasis-open.org/mqtt/mqtt/v5.0/mqtt-v5.0.html}
\end{botherref}
\endbibitem

\bibitem{mosquitto:2017}
\begin{barticle}
\bauthor{\bsnm{Light}, \binits{R.A.}}:
\batitle{{Mosquitto: server and client implementation of the MQTT protocol}}.
\bjtitle{Journal of Open Source Software}
\bvolume{2}(\bissue{13}),
\bfpage{265}
(\byear{2017})
\end{barticle}
\endbibitem

\bibitem{netem:2011}
\begin{botherref}
\oauthor{\bsnm{Hemminger}, \binits{S.}},
\oauthor{\bsnm{Ludovici}, \binits{F.}},
\oauthor{\bsnm{Pfeifer}, \binits{H.P.}}:
{NetEm - Network Emulator}.
Accessed: 2020-08-22
(2011).
\url{https://man7.org/linux/man-pages/man8/tc-netem.8.html}
\end{botherref}
\endbibitem

\end{thebibliography}

\newcommand{\BMCxmlcomment}[1]{}

\BMCxmlcomment{

<refgrp>

<bibl id="B1">
  <title><p>{Origins of the Digital Twin Concept}</p></title>
  <aug>
    <au><snm>Grieves</snm><fnm>M</fnm></au>
  </aug>
  <pubdate>2016</pubdate>
</bibl>

<bibl id="B2">
  <title><p>Real-time systems: design principles for distributed embedded
  applications</p></title>
  <aug>
    <au><snm>Kopetz</snm><fnm>H</fnm></au>
  </aug>
  <publisher>Wien, AT: Springer Science \& Business Media</publisher>
  <pubdate>2011</pubdate>
</bibl>

<bibl id="B3">
  <title><p>Digital Twin: Mitigating Unpredictable, Undesirable Emergent
  Behavior in Complex Systems</p></title>
  <aug>
    <au><snm>Grieves</snm><fnm>M</fnm></au>
    <au><snm>Vickers</snm><fnm>J</fnm></au>
  </aug>
  <source>Transdisciplinary Perspectives on Complex Systems: New Findings and
  Approaches</source>
  <publisher>Rochester, USA: Springer International Publishing</publisher>
  <editor>Kahlen, Franz-Josef and Flumerfelt, Shannon and Alves,
  Anabela</editor>
  <edition>1</edition>
  <pubdate>2017</pubdate>
  <fpage>85</fpage>
  <lpage>-113</lpage>
</bibl>

<bibl id="B4">
  <title><p>{B-space: dynamic management and assurance of open systems of
  systems}</p></title>
  <aug>
    <au><snm>Schneider</snm><fnm>D</fnm></au>
    <au><snm>Trapp</snm><fnm>M</fnm></au>
  </aug>
  <source>{Journal of Internet Services and Applications}</source>
  <publisher>SpringerOpen</publisher>
  <pubdate>2018</pubdate>
  <volume>9</volume>
  <issue>1</issue>
  <fpage>1</fpage>
  <lpage>-16</lpage>
</bibl>

<bibl id="B5">
  <title><p>All one needs to know about fog computing and related edge
  computing paradigms: A complete survey</p></title>
  <aug>
    <au><snm>Yousefpour</snm><fnm>A</fnm></au>
    <au><snm>Fung</snm><fnm>C</fnm></au>
    <au><snm>Nguyen</snm><fnm>T</fnm></au>
    <au><snm>Kadiyala</snm><fnm>K</fnm></au>
    <au><snm>Jalali</snm><fnm>F</fnm></au>
    <au><snm>Niakanlahiji</snm><fnm>A</fnm></au>
    <au><snm>Kong</snm><fnm>J</fnm></au>
    <au><snm>Jue</snm><fnm>JP</fnm></au>
  </aug>
  <source>Journal of Systems Architecture</source>
  <pubdate>2019</pubdate>
  <volume>98</volume>
  <fpage>289</fpage>
  <lpage>330</lpage>
  <url>http://www.sciencedirect.com/science/article/pii/S1383762118306349</url>
</bibl>

<bibl id="B6">
  <title><p>Edge Computing: Vision and Challenges</p></title>
  <aug>
    <au><snm>{Shi}</snm><fnm>W.</fnm></au>
    <au><snm>{Cao}</snm><fnm>J.</fnm></au>
    <au><snm>{Zhang}</snm><fnm>Q.</fnm></au>
    <au><snm>{Li}</snm><fnm>Y.</fnm></au>
    <au><snm>{Xu}</snm><fnm>L.</fnm></au>
  </aug>
  <source>IEEE Internet of Things Journal</source>
  <pubdate>2016</pubdate>
  <volume>3</volume>
  <issue>5</issue>
  <fpage>637</fpage>
  <lpage>646</lpage>
</bibl>

<bibl id="B7">
  <title><p>Industrial internet of things: Recent advances, enabling
  technologies and open challenges</p></title>
  <aug>
    <au><snm>Khan</snm><fnm>WZ</fnm></au>
    <au><snm>Rehman</snm><fnm>MH</fnm></au>
    <au><snm>Zangoti</snm><fnm>HM</fnm></au>
    <au><snm>Afzal</snm><fnm>MK</fnm></au>
    <au><snm>Armi</snm><fnm>N</fnm></au>
    <au><snm>Salah</snm><fnm>K</fnm></au>
  </aug>
  <source>Computers \& Electrical Engineering</source>
  <publisher>Elsevier</publisher>
  <pubdate>2020</pubdate>
  <volume>81</volume>
  <fpage>106522</fpage>
</bibl>

<bibl id="B8">
  <title><p>Fog computing for energy-aware load balancing and scheduling in
  smart factory</p></title>
  <aug>
    <au><snm>Wan</snm><fnm>J</fnm></au>
    <au><snm>Chen</snm><fnm>B</fnm></au>
    <au><snm>Wang</snm><fnm>S</fnm></au>
    <au><snm>Xia</snm><fnm>M</fnm></au>
    <au><snm>Li</snm><fnm>D</fnm></au>
    <au><snm>Liu</snm><fnm>C</fnm></au>
  </aug>
  <source>IEEE Transactions on Industrial Informatics</source>
  <publisher>IEEE</publisher>
  <pubdate>2018</pubdate>
  <volume>14</volume>
  <issue>10</issue>
  <fpage>4548</fpage>
  <lpage>-4556</lpage>
</bibl>

<bibl id="B9">
  <title><p>{Fog assisted-IoT enabled patient health monitoring in smart
  homes}</p></title>
  <aug>
    <au><snm>Verma</snm><fnm>P</fnm></au>
    <au><snm>Sood</snm><fnm>SK</fnm></au>
  </aug>
  <source>IEEE Internet of Things Journal</source>
  <publisher>IEEE</publisher>
  <pubdate>2018</pubdate>
  <volume>5</volume>
  <issue>3</issue>
  <fpage>1789</fpage>
  <lpage>-1796</lpage>
</bibl>

<bibl id="B10">
  <title><p>A hierarchical distributed fog computing architecture for big data
  analysis in smart cities</p></title>
  <aug>
    <au><snm>Tang</snm><fnm>B</fnm></au>
    <au><snm>Chen</snm><fnm>Z</fnm></au>
    <au><snm>Hefferman</snm><fnm>G</fnm></au>
    <au><snm>Wei</snm><fnm>T</fnm></au>
    <au><snm>He</snm><fnm>H</fnm></au>
    <au><snm>Yang</snm><fnm>Q</fnm></au>
  </aug>
  <source>Proceedings of the ASE BigData \& SocialInformatics 2015</source>
  <pubdate>2015</pubdate>
  <fpage>1</fpage>
  <lpage>-6</lpage>
</bibl>

<bibl id="B11">
  <title><p>Fog computing for sustainable smart cities: A survey</p></title>
  <aug>
    <au><snm>Perera</snm><fnm>C</fnm></au>
    <au><snm>Qin</snm><fnm>Y</fnm></au>
    <au><snm>Estrella</snm><fnm>JC</fnm></au>
    <au><snm>Reiff Marganiec</snm><fnm>S</fnm></au>
    <au><snm>Vasilakos</snm><fnm>AV</fnm></au>
  </aug>
  <source>ACM Computing Surveys (CSUR)</source>
  <publisher>ACM New York, NY, USA</publisher>
  <pubdate>2017</pubdate>
  <volume>50</volume>
  <issue>3</issue>
  <fpage>1</fpage>
  <lpage>-43</lpage>
</bibl>

<bibl id="B12">
  <title><p>{An NFV-powered emergency system for smart enhanced living
  environments}</p></title>
  <aug>
    <au><snm>Nikoloudakis</snm><fnm>Y</fnm></au>
    <au><snm>Markakis</snm><fnm>E</fnm></au>
    <au><snm>Mastorakis</snm><fnm>G</fnm></au>
    <au><snm>Pallis</snm><fnm>E</fnm></au>
    <au><snm>Skianis</snm><fnm>C</fnm></au>
  </aug>
  <source>2017 IEEE Conference on Network Function Virtualization and Software
  Defined Networks (NFV-SDN)</source>
  <pubdate>2017</pubdate>
  <fpage>258</fpage>
  <lpage>-263</lpage>
</bibl>

<bibl id="B13">
  <title><p>Edge-oriented computing paradigms: A survey on architecture design
  and system management</p></title>
  <aug>
    <au><snm>Li</snm><fnm>C</fnm></au>
    <au><snm>Xue</snm><fnm>Y</fnm></au>
    <au><snm>Wang</snm><fnm>J</fnm></au>
    <au><snm>Zhang</snm><fnm>W</fnm></au>
    <au><snm>Li</snm><fnm>T</fnm></au>
  </aug>
  <source>ACM Computing Surveys (CSUR)</source>
  <publisher>ACM New York, NY, USA</publisher>
  <pubdate>2018</pubdate>
  <volume>51</volume>
  <issue>2</issue>
  <fpage>1</fpage>
  <lpage>-34</lpage>
</bibl>

<bibl id="B14">
  <title><p>A Survey of Communication Protocols for Internet of Things and
  Related Challenges of Fog and Cloud Computing Integration</p></title>
  <aug>
    <au><snm>Dizdarevi\'{c}</snm><fnm>J</fnm></au>
    <au><snm>Carpio</snm><fnm>F</fnm></au>
    <au><snm>Jukan</snm><fnm>A</fnm></au>
    <au><snm>Masip Bruin</snm><fnm>X</fnm></au>
  </aug>
  <source>ACM Comput. Surv.</source>
  <publisher>New York, NY, USA: Association for Computing Machinery</publisher>
  <pubdate>2019</pubdate>
  <volume>51</volume>
  <issue>6</issue>
</bibl>

<bibl id="B15">
  <title><p>{How Digital Twins Enable the Next Level of PLM – A Guide for the
  Concept and the Implementation in the Internet of Everything Era}</p></title>
  <aug>
    <au><snm>Voell</snm><fnm>C</fnm></au>
    <au><snm>Chatterjee</snm><fnm>P</fnm></au>
    <au><snm>Rauch</snm><fnm>A</fnm></au>
    <au><snm>Golovatchev</snm><fnm>J</fnm></au>
  </aug>
  <source>Product Lifecycle Management to Support Industry 4.0</source>
  <publisher>Cham, CH: Springer International Publishing</publisher>
  <pubdate>2018</pubdate>
  <fpage>238</fpage>
  <lpage>249</lpage>
</bibl>

<bibl id="B16">
  <title><p>{Data-driven Digital Twin approach for process optimization: an
  industry use case}</p></title>
  <aug>
    <au><snm>Stojanovic</snm><fnm>N</fnm></au>
    <au><snm>Milenovic</snm><fnm>D</fnm></au>
  </aug>
  <source>2018 IEEE International Conference on Big Data</source>
  <pubdate>2018</pubdate>
</bibl>

<bibl id="B17">
  <title><p>{Classification of cyber-physical production systems applications:
  Proposition of an analysis framework}</p></title>
  <aug>
    <au><snm>Cardin</snm><fnm>O</fnm></au>
  </aug>
  <source>Computers in Industry</source>
  <publisher>Elsevier</publisher>
  <pubdate>2019</pubdate>
  <volume>104</volume>
  <fpage>11</fpage>
  <lpage>-21</lpage>
</bibl>

<bibl id="B18">
  <title><p>{The NIST definition of cloud computing}</p></title>
  <aug>
    <au><snm>Mell</snm><fnm>P</fnm></au>
    <au><snm>Grance</snm><fnm>T</fnm></au>
    <au><cnm>others</cnm></au>
  </aug>
  <source>National Institute of Science and Technology, Special Publication,
  800</source>
  <publisher>Computer Security Division, Information Technology Laboratory,
  National Institute of Standards and Technology</publisher>
  <pubdate>2011</pubdate>
  <volume>145</volume>
</bibl>

<bibl id="B19">
  <title><p>Cloud computing: theory and practice</p></title>
  <aug>
    <au><snm>Marinescu</snm><fnm>DC</fnm></au>
  </aug>
  <publisher>Waltham, USA: Morgan Kaufmann</publisher>
  <pubdate>2017</pubdate>
</bibl>

<bibl id="B20">
  <title><p>Fault-tolerant and reliable computation in cloud
  computing</p></title>
  <aug>
    <au><snm>Deng</snm><fnm>J</fnm></au>
    <au><snm>Huang</snm><fnm>SCH</fnm></au>
    <au><snm>Han</snm><fnm>YS</fnm></au>
    <au><snm>Deng</snm><fnm>JH</fnm></au>
  </aug>
  <source>2010 IEEE Globecom Workshops</source>
  <pubdate>2010</pubdate>
  <fpage>1601</fpage>
  <lpage>-1605</lpage>
</bibl>

<bibl id="B21">
  <title><p>What is a Fog Node? A Tutorial on Current Concepts towards a Common
  Definition</p></title>
  <aug>
    <au><snm>Tordera</snm><fnm>EM</fnm></au>
    <au><snm>Masip Bruin</snm><fnm>X</fnm></au>
    <au><snm>Garcia Alminana</snm><fnm>J</fnm></au>
    <au><snm>Jukan</snm><fnm>A</fnm></au>
    <au><snm>Ren</snm><fnm>GJ</fnm></au>
    <au><snm>Zhu</snm><fnm>J</fnm></au>
    <au><snm>Farr{\'e}</snm><fnm>J</fnm></au>
  </aug>
  <source>arXiv preprint arXiv:1611.09193</source>
  <pubdate>2016</pubdate>
</bibl>

<bibl id="B22">
  <title><p>On blockchain and its integration with IoT. Challenges and
  opportunities</p></title>
  <aug>
    <au><snm>Reyna</snm><fnm>A</fnm></au>
    <au><snm>Martín</snm><fnm>C</fnm></au>
    <au><snm>Chen</snm><fnm>J</fnm></au>
    <au><snm>Soler</snm><fnm>E</fnm></au>
    <au><snm>Díaz</snm><fnm>M</fnm></au>
  </aug>
  <source>Future Generation Computer Systems</source>
  <pubdate>2018</pubdate>
  <volume>88</volume>
  <fpage>173</fpage>
  <lpage>190</lpage>
  <url>https://www.sciencedirect.com/science/article/pii/S0167739X17329205</url>
</bibl>

<bibl id="B23">
  <title><p>Deploying data-intensive applications with multiple services
  components on edge</p></title>
  <aug>
    <au><snm>Chen</snm><fnm>Y</fnm></au>
    <au><snm>Deng</snm><fnm>S</fnm></au>
    <au><snm>Ma</snm><fnm>H</fnm></au>
    <au><snm>Yin</snm><fnm>J</fnm></au>
  </aug>
  <source>Mobile Networks and Applications</source>
  <publisher>Springer</publisher>
  <pubdate>2019</pubdate>
  <fpage>1</fpage>
  <lpage>-16</lpage>
</bibl>

<bibl id="B24">
  <title><p>{Fog Computing and the Internet of Things: A Review}</p></title>
  <aug>
    <au><snm>Atlam</snm><fnm>HF</fnm></au>
    <au><snm>Walters</snm><fnm>RJ</fnm></au>
    <au><snm>Wills</snm><fnm>GB</fnm></au>
  </aug>
  <source>Big Data and Cognitive Computing</source>
  <publisher>Multidisciplinary Digital Publishing Institute</publisher>
  <pubdate>2018</pubdate>
  <volume>2</volume>
  <issue>2</issue>
  <fpage>10</fpage>
</bibl>

<bibl id="B25">
  <title><p>{Edge computing for Internet of Things: A survey, e-healthcare case
  study and future direction}</p></title>
  <aug>
    <au><snm>Ray</snm><fnm>PP</fnm></au>
    <au><snm>Dash</snm><fnm>D</fnm></au>
    <au><snm>De</snm><fnm>D</fnm></au>
  </aug>
  <source>{Journal of Network and Computer Applications}</source>
  <pubdate>2019</pubdate>
  <volume>140</volume>
  <fpage>1</fpage>
  <lpage>22</lpage>
</bibl>

<bibl id="B26">
  <title><p>Offloading in fog computing for IoT: Review, enabling technologies,
  and research opportunities</p></title>
  <aug>
    <au><snm>Aazam</snm><fnm>M</fnm></au>
    <au><snm>Zeadally</snm><fnm>S</fnm></au>
    <au><snm>Harras</snm><fnm>KA</fnm></au>
  </aug>
  <source>Future Generation Computer Systems</source>
  <pubdate>2018</pubdate>
  <volume>87</volume>
  <fpage>278</fpage>
  <lpage>289</lpage>
</bibl>

<bibl id="B27">
  <title><p>{Modeling of Cyber-Physical Systems and Digital Twin Based on Edge
  Computing, Fog Computing and Cloud Computing Towards Smart
  Manufacturing}</p></title>
  <aug>
    <au><snm>Qi</snm><fnm>Q</fnm></au>
    <au><snm>Zhao</snm><fnm>D</fnm></au>
    <au><snm>Liao</snm><fnm>TW</fnm></au>
    <au><snm>Tao</snm><fnm>F</fnm></au>
  </aug>
  <source>ASME 2018 13th International Manufacturing Science and Engineering
  Conference</source>
  <pubdate>2018</pubdate>
</bibl>

<bibl id="B28">
  <title><p>{Cyber Physical Systems Framework of Edge-, Fog-, and
  Cloud-Computing}</p></title>
  <aug>
    <au><snm>Kim</snm><fnm>T</fnm></au>
  </aug>
  <source>Proceedings of the International Conference on Embedded Systems,
  Cyber-physical Systems, and Applications (ESCS)</source>
  <pubdate>2019</pubdate>
  <fpage>10</fpage>
  <lpage>-14</lpage>
</bibl>

<bibl id="B29">
  <title><p>{Edge-Fog cloud: A distributed cloud for Internet of Things
  computations}</p></title>
  <aug>
    <au><snm>Mohan</snm><fnm>N</fnm></au>
    <au><snm>Kangasharju</snm><fnm>J</fnm></au>
  </aug>
  <source>2016 Cloudification of the Internet of Things (CIoT)</source>
  <pubdate>2016</pubdate>
  <fpage>1</fpage>
  <lpage>-6</lpage>
</bibl>

<bibl id="B30">
  <title><p>{Deployment of Fog and Edge Computing in IoT for Cyber-Physical
  Infrastructures in the 5G Era}</p></title>
  <aug>
    <au><snm>Ahmad</snm><fnm>S</fnm></au>
    <au><snm>Afzal</snm><fnm>MM</fnm></au>
  </aug>
  <source>International Conference on Sustainable Communication Networks and
  Application</source>
  <pubdate>2019</pubdate>
  <fpage>351</fpage>
  <lpage>-359</lpage>
</bibl>

<bibl id="B31">
  <title><p>{A Fog Computing Based Cyber-Physical System for the Automation of
  Pipe-Related Tasks in the Industry 4.0 Shipyard}</p></title>
  <aug>
    <au><snm>Fern{\'a}ndez Caram{\'e}s</snm><fnm>TM</fnm></au>
    <au><snm>Fraga Lamas</snm><fnm>P</fnm></au>
    <au><snm>Su{\'a}rez Albela</snm><fnm>M</fnm></au>
    <au><snm>D{\'\i}az Bouza</snm><fnm>MA</fnm></au>
  </aug>
  <source>Sensors</source>
  <publisher>Multidisciplinary Digital Publishing Institute</publisher>
  <pubdate>2018</pubdate>
  <volume>18</volume>
  <issue>6</issue>
  <fpage>1961</fpage>
</bibl>

<bibl id="B32">
  <title><p>An open source approach to the design and implementation of Digital
  Twins for Smart Manufacturing</p></title>
  <aug>
    <au><snm>Damjanovic Behrendt</snm><fnm>V</fnm></au>
    <au><snm>Behrendt</snm><fnm>W</fnm></au>
  </aug>
  <source>International Journal of Computer Integrated Manufacturing</source>
  <publisher>Taylor & Francis</publisher>
  <pubdate>2019</pubdate>
  <volume>32</volume>
  <issue>4-5</issue>
  <fpage>366</fpage>
  <lpage>384</lpage>
</bibl>

<bibl id="B33">
  <title><p>{MQTT Version 5.0}</p></title>
  <aug>
    <au><snm>Banks</snm><fnm>A</fnm></au>
    <au><snm>Briggs</snm><fnm>E</fnm></au>
    <au><snm>Borgendale</snm><fnm>K</fnm></au>
    <au><snm>Gupta</snm><fnm>R</fnm></au>
  </aug>
  <source>OASIS Standard</source>
  <pubdate>2019</pubdate>
  <url>https://docs.oasis-open.org/mqtt/mqtt/v5.0/mqtt-v5.0.html</url>
  <note>Accessed: 2020-08-09</note>
</bibl>

<bibl id="B34">
  <title><p>{Mosquitto: server and client implementation of the MQTT
  protocol}</p></title>
  <aug>
    <au><snm>Light</snm><fnm>RA</fnm></au>
  </aug>
  <source>Journal of Open Source Software</source>
  <publisher>The Open Journal</publisher>
  <pubdate>2017</pubdate>
  <volume>2</volume>
  <issue>13</issue>
  <fpage>265</fpage>
  <url>https://doi.org/10.21105/joss.00265</url>
</bibl>

<bibl id="B35">
  <title><p>{NetEm - Network Emulator}</p></title>
  <aug>
    <au><snm>Hemminger</snm><fnm>S</fnm></au>
    <au><snm>Ludovici</snm><fnm>F</fnm></au>
    <au><snm>Pfeifer</snm><fnm>HP</fnm></au>
  </aug>
  <pubdate>2011</pubdate>
  <url>https://man7.org/linux/man-pages/man8/tc-netem.8.html</url>
  <note>Accessed: 2020-08-22</note>
</bibl>

</refgrp>
} 


\end{backmatter}
\end{document}